\documentclass{aa}
\usepackage[varg]{txfonts}
\usepackage{booktabs,graphicx,siunitx,xspace,natbib}
\bibpunct{(}{)}{;}{a}{}{,} 
\DeclareSIUnit \parsec {pc}
\DeclareSIUnit \pc     {\parsec}
\DeclareSIUnit \kpc    {\kilo \parsec}
\DeclareSIUnit \msun   {\ensuremath{M_\sun}}
\DeclareSIUnit \hour   {h}
\DeclareSIUnit \year   {a}
\DeclareSIUnit \keV    {keV}
\DeclareSIUnit \km     {km}
\DeclareSIUnit \s      {s}

\newcommand{\about}{$\simeq$}

\newcommand{\Fe}{$^{60}$Fe\xspace}
\newcommand{\Co}{$^{56}$Co\xspace} 

\newcommand{\Ni}{$^{56}$Ni\xspace}
\newcommand{\Ti}{$^{44}$Ti\xspace}
\newcommand{\Sc}{$^{44}$Sc\xspace}
\newcommand{\Ca}{$^{44}$Ca\xspace}
\newcommand{\Msol}{M\ensuremath{_\odot}\xspace}

\newcommand{\cms}{cm\ensuremath{^{-2}} s\ensuremath{^{-1}}\xspace}

\usepackage{color}

\begin{document}

\title{Revisiting INTEGRAL/SPI observations of $^{44}$Ti from Cassiopeia A}

\author{
  Thomas Siegert   \inst{\ref{inst:mpe}}\thanks{E-mail: tsiegert@mpe.mpg.de} \and
  Roland Diehl         \inst{\ref{inst:mpe}, \ref{inst:xcu}} \and
  Martin G.~H. Krause        \inst{\ref{inst:xcu}, \ref{inst:mpe}, \ref{inst:usm}} \and
  Jochen Greiner       \inst{\ref{inst:mpe}, \ref{inst:xcu}}
}

\institute{
  Max-Planck-Institut f\"ur extraterrestrische Physik,
  D-85741 Garching, Germany
  \label{inst:mpe}
  \and
  Excellence-Cluster ``Origin \& Structure of Universe'',
  D-85748 Garching, Germany
  \label{inst:xcu}
  \and
  Universit\"ats-Sternwarte Ludwig-Maximilians-Universit\"at,
  D-81679 M\"unchen, Germany
  \label{inst:usm}
}

\date{Received February 12, 2015; accepted May 20, 2015}

\abstract
{The 340-year old supernova remnant Cassiopeia A, located at 3.4 kpc distance, is the best-studied young core-collapse supernova remnant. Nucleosynthesis yields in radioactive isotopes have been studied with different methods, in particular, methods for 
  production and ejection of $^{44}$Ti and $^{56}$Ni, which originate in the innermost regions of the supernova.
 \Ti was first discovered in this remnant, but is not seen consistently in other core-collapse sources.
 }
{We aim to measure radioactive \Ti ejected in Cassiopeia A and to place constraints on velocities of these ejecta by determining X- and $\gamma$-ray line-shape parameters of the emission lines.}
{We analyzed the observations made with the SPI spectrometer on INTEGRAL
  together with an improved instrumental background method, to
  achieve a high spectroscopic resolution that enables interpretation for a velocity constraint
  on \Ti ejecta from the 1.157~MeV $\gamma$-ray line of the \Sc decay.}
{We observe both the hard X-ray line at 78~keV and the $\gamma$-ray line at 1157~keV from the \Ti decay chain
at a combined significance of 3.8~$\sigma$.
Measured fluxes are $(2.1 \pm 0.4)~10^{-5}$~ph~\cms and $(3.5 \pm 1.2)~10^{-5}$~ph~\cms, which corresponds to $(1.5\pm0.4)~10^{-4}$ and $(2.4\pm0.9)~10^{-4}$~\Msol of \Ti, respectively.
The measured Doppler broadening of the lines implies expansion velocities of 4300 and 2200~$\mathrm{km~s^{-1}}$, respectively. 
 By combining our results with previous studies, we determine a more precise estimate of ejected \Ti of $(1.37\pm0.19)~10^{-4}$~\Msol.}
 {The measurements of the two lines are consistent with previous studies. The flux in the line originating from excited $^{44}$Ca is significantly higher than the flux determined in the lines from $^{44}$Sc. Cosmic-ray acceleration within the supernova remnant may be responsible for an additional contribution to this line from nuclear de-excitation following energetic particle collisions in the remnant and swept-up material. }

\keywords{
 (Stars:) supernovae: individual: Cas A --
  Nuclear reactions, nucleosynthesis, abundances --
  Stars: massive --
  ISM: supernova remnants --
  Gamma rays: ISM --
  Techniques: spectroscopic
}

\maketitle

%
\section{Introduction}
\label{sec:intro}
Cassiopeia A (Cas A) is the closest young remnant of a core-collapse supernova and therefore a prominent study object in many wavelength regimes. It may have been seen in AD 1680 by Flamsted~\citep{Ashworth1980_CasA,Green2002_CasA}, but was not as bright as expected from its proximity of 3.4$^{+0.3}_{-0.1}$~kpc~\citep{Reed1995_CasA}. Apparently, the supernova itself was occulted by interstellar dust and gas, and its brightness and \Ni production are therefore only inferred indirectly~\citep{Eriksen2009_CasA}. The date of the event is not precisely settled; \citet{Thorstensen2001_CasA} inferred the supernova event at AD 1671.3$\pm$0.9 by measuring the proper motion of the supernova remnant (SNR) bright main shell as well as faint, high-velocity, outer ejecta knots, and assuming free expansion.

Cas A was the first supernova in which the characteristic line at 1157~keV from \Ti decay had been reported~\citep{Iyudin1994_CasA}; this line arises from de-excitation of $^{44}$Ca, the final daughter nucleus in the decay chain $^{44}$Ti$\rightarrow$$^{44}$Sc$\rightarrow$$^{44}$Ca. Later measurements focused on the equally bright lines from the first stage in the \Ti decay chain, the de-excitation of $^{44}$Sc. In particular, the studies with INTEGRAL/IBIS~\citep{Renaud2006_CasA} and NuStar~\citep{Grefenstette2014_CasA} settled the amount of \Ti to \about~1.4~10$^{-4}$~\Msol (IBIS: 1.6$^{+0.6}_{-0.3}$~10$^{-4}$~\Msol; NuStar: (1.25$\pm$0.3)~10$^{-4}$~\Msol). This is significantly higher than the yields predicted for typical supernova models, which suggested that Cas A represents a rare subclass of \Ti ejecting supernovae, and explosion asymmetries have been thought to cause such anomalies~\citep{Nagataki1998_CasA,The2006_44Ti}. The \Ti image recently obtained by the NuStar hard X-ray telescope~\citep{Grefenstette2014_CasA} has shown that Cas A ejecta carrying \Ti appear in clumps, re-affirming that Cas A did not explode as a spherically symmetric supernova. Other observations had shown this before from X-ray measurements,
for example, and especially from optical/IR light echo spectra, revealing this asymmetry in the supernova photosphere. Furthermore, it was shown by these light echoes that Cas A exploded as a type IIb supernova~\citep[][see also discussion by~\citet{Wheeler2008_CasA}]{Krause2005_CasA,Krause2008_CasA,Rest2008_CasA,Rest2011_CasA}.

Earlier gamma-ray spectroscopy with INTEGRAL/SPI had provided some velocity constraints on the radioactive ejecta from the inner supernova, with a lower limit on \Ti ejecta of 500~km~s$^{-1}$ from a nondetection of the high-energy line at 1.157 MeV, which was attributed to Doppler broadening~\citep{Martin2007_CasA}. Doppler broadening is proportional to photon energy; therefore, a background dominated measurement such as with SPI would be more sensitive to a low-energy line, while the high-energy line would drown in the background, especially if part of the \Ti ejecta were moving at high velocities. NuStar detectors also provide a sufficiently high spectral resolution for velocity information. \citet{Grefenstette2014_CasA} reported an overall velocity of fastest \Ti clumps of $(5350\pm1610)$~km~s$^{-1}$, simultaneous to a bulk line-of-sight velocity in the range of 2000~km~s$^{-1}$. The NuStar result is only based on the detection of the 68~keV line, although both hard X-ray lines have been measured. Our recently improved method for handling instrumental background suggested re-analysis of INTEGRAL data from Cas A with the aim to revisit and refine velocity constraints from high-resolution spectroscopy with the SPI instrument of all three lines resulting from the \Ti decay chain.

\Ti and \Ni arise from nuclear burning in the innermost parts of supernovae very close to the mass-cut separating ejecta from the compact remnant~\citep{Diehl1998_ns}. Therefore \Ti and \Ni are expected to also be co-spatial when they decay in the expanding SNR~\citep{The1998_44Ti,Hoffman2010_44Ti,Magkotsios2010_44Ti}. But as this inner region is probably characterized by dynamical instabilities and simultaneous inflows and outflows of material, it remains unclear how and where this separation between material accreting onto the compact remnant and the ejecta occurs, and thus how much of the \Ti can end up in the ejecta~\citep{Fryer2008_SN,Wongwathanarat2013_SN,Popov2014_SN}.

\Ni radioactive decay occurs with a first decay to \Co after $\tau$~\about~8 days~\citep{daCruz1996_56Ni}, that is, when supernovae are still sufficiently dense to absorb even gamma-rays at MeV energies and convert this radioactivity energy into thermal emission, which makes supernovae shine at UV to IR wavelengths~\citep{Isern2008_SNIa,Roepke2012_SNIa}. As the supernova envelope becomes more transparent, fewer gamma-rays thermalize, in particular when the second decay stage from \Co to \Fe at $\tau$~\about~111 days~\citep{Funck1992_56Co} produces 0.847 and 1.238~MeV gamma-rays. Therefore, these characteristic gamma-rays are expected to be observable with gamma-ray telescopes a few months after the explosion~\citep{Isern2008_SNIa}. \Ti has a considerably longer decay lifetime, and thus gamma-rays will escape readily. \Ti decays to \Sc within $\tau$ \about~86~years \citep{Ahmad1998_44Ti,Goerres1998_44Ti,Norman1998_44Ti,Wietfeldt1999_44Ti,Hashimoto2001_44Ti,Ahmad2006_44Ti}. In this first decay stage,  gamma-rays of 67.87 and 78.36 keV are emitted. The subsequent decay of \Sc to \Ca occurs after $\tau$~\about~5.73~hours~\citep{Audi2003_halflife}, producing a characteristic gamma-ray line at 1157.02 keV. But in spite of this long decay time, significant deposit of radioactivity energy is expected in the late SNR from positrons emitted during these $\beta^+$~decays.

For SN1987A, this has been investigated: The study of the late light-curve in SN1987A~\citep{Lundqvist2001_SN87A,Fransson2002_SN87A,Jerkstrand2011_SN87A} obtained an inferred \Ti mass from optical and infrared spectra of SN1987A to (1.5$\pm$0.5)~10$^{-4}$~\Msol. The \Ti that mainly powers the late light-curve (years after the explosion) has a direct effect on the total absolute flux level of the optical-to-infrared spectra, but also depends on the estimated extinction. Corresponding gamma-ray emission has recently been detected by~\citet{Grebenev2012_SN1987A} with IBIS; this direct measurement constrained the \Ti mass of SN1987A to (3.1$\pm$0.8)~10$^{-4}$~\Msol (see, however,~\citet{Seitenzahl2014_SN87A}, who obtained (0.55$\pm$0.17)~10$^{-4}$~\Msol from a detailed, multi-component light-curve model). Beyond SN1987A, the Cas A SNR is the only other core-collapse event where this consistency between nucleosynthesis radioactive elements and observed emission can be studied.

%
\section{Data and analysis}
\label{sec:spi-analysis}
\subsection{Instrument, exposure, and general analysis method}
\label{subsec:treatment}

%

\begin{figure}
  \centering
  \includegraphics[width=\linewidth]{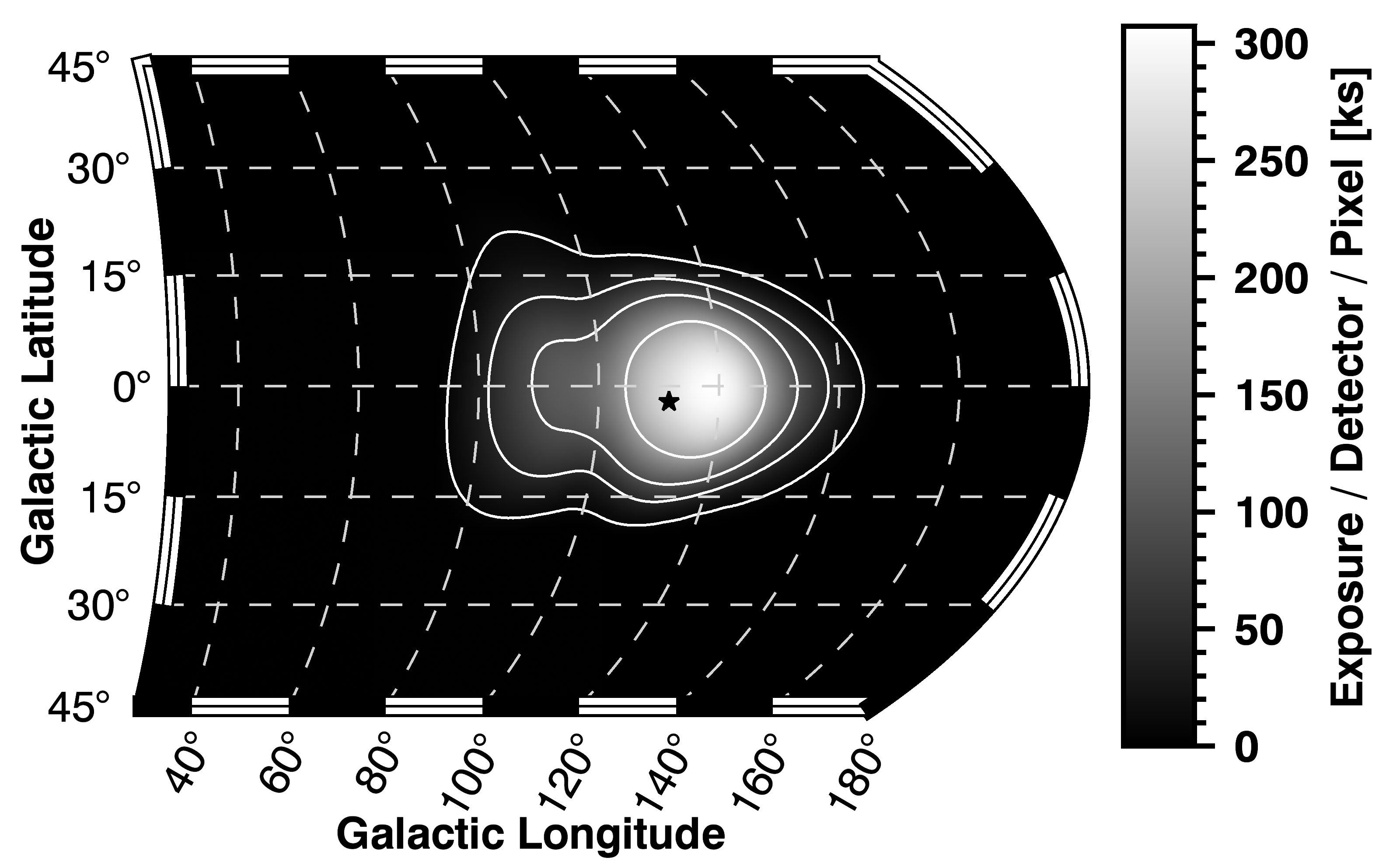}
  \caption{Exposure map of the sky with the SPI telescope on INTEGRAL for the data used in this analysis (Mar 2003 -- Jan 2014). Cas A is marked with a star at its position $(l/b) = (111.735^{\circ}/-2.130^{\circ})$. The total exposure time in the PCFOV is 10.84 Ms, while Cas A was seen fully coded within an exposure of 5.15 Ms. }
  \label{fig:spimodfit-sa-exposuremap}
\end{figure}

The INTEGRAL space gamma-ray observatory~\citep{Winkler2003_INTEGRAL} carries the spectrometer instrument SPI as one of its two main instruments, measuring gamma-rays in the energy range of 20~keV to 8~MeV with a spectral resolution of $\sim2.3$~keV at 1~MeV~\citep{Vedrenne2003_SPI,Roques2003_SPI}. SPI features a camera consisting of 19 Ge detectors with high
spectral resolution, which measures celestial gamma-rays through shadograms imprinted by a tungsten coded mask 171~cm above the Ge camera. A high instrumental background from cosmic-ray activation of spacecraft and instrument materials makes it challenging to distinguish these shadowgrams from the total signal. 
SPI data consist of energy-binned spectra for each of the 19 Ge detectors of the telescope camera, typically taken
in \SI{30}{\minute} exposures dithered in $\sim$2$^{\circ}$ steps in a certain sky region around the target of interest. 
For our analysis, we used exposures in which Cas A was at least in the partially coded field of view (PCFOV) of SPI of $34^{\circ}\times34^{\circ}$ accumulated over almost eleven years of the INTEGRAL mission. Solar flares mainly affect the background rate and are difficult to model. We therefore excluded data of any orbit in which a solar flare occurred until the overall background returned to normal. Perigee passages around Earth, involving a higher radiation dose when the satellite is passing the Van Allen radiation belts, additionally affect the background rate, and we consequently chose data from orbital phases $\left[0.15,0.85\right]$. Background variations due to cooling (heating) after (before) annealing periods were removed accordingly. Finally, our 11-year observation set consists of \num{5604} instrument pointings, resulting in a total exposure time of 10.84~Ms (PCFOV) in which Cas A was seen fully coded by the SPI coded mask aperture during 5.15~Ms (see Fig.~\ref{fig:spimodfit-sa-exposuremap}). 

In our spectroscopic analysis we generally fit scaling factors of an adopted model of the sky intensity distribution plus a model of instrumental
background to the set of spectra accumulated during the multiyear observations from our 19 Ge detectors. This resembles linear regression 
of specific, given time sequences, incorporating the instrumental (imaging) response function (IRF) for the sky contribution to the data. Multiple components of sky and background can be used as required by adequacy of the fit (mainly background) and the scientific question (sky components).
Hence, data $d_k$ of energy bin $k$ are modeled as a linear combination of the $N_I$ sky model
components $M_{ij}$ as convolved with the instrument response matrix $R_{jk}$, and the $N_B$ background components $B_{jk}$:
\begin{equation}
d_k = \sum_j R_{jk} \sum_{i = 1}^{N_\mathrm{I}} \theta_i M_{ij} +
\sum_{i = N_\mathrm{I} + 1}^{N_\mathrm{I} + N_\mathrm{B}} \theta_i
B_{jk}\label{eq:model-fit}
.\end{equation}
The comparison between scaled models and measurements is performed in data space, which consists of the
counts per energy bin $k$ measured in each of the SPI detectors $j$ for each
single telescope pointing as part of the complete observation. 
Our adopted sky intensity distribution for Cas A is a point source, given the SPI imaging resolution of 2.7 degrees and the total SNR diameter of $\sim$5~arcmin of Cas A. The model for the behavior of the instrumental background is derived from a separate analysis using all available data during and around the time of Cas A exposure. We repeat this fit (model scaling) for \SI{0.5}{\keV} wide energy intervals to obtain the source intensity spectrum, that is, without assuming a prior spectral shape. 

A priori, we do not assume the processes that lead to the low-energy
lines (68~keV and 78~keV) that originate in excited \Sc, to be the same as for the high-energy line (1157~keV) from excited \Ca. For example, cosmic-ray (CR) collisions with ambient \Ca at several tens of MeV/nucleon effectively produce nuclear excitation, followed by de-excitation photons at 1.157~MeV. Because $^{44}$Sc is short lived, no corresponding excitation of $^{44}$Sc will occur through this cosmic-ray process, meaning that no additional emission is produced in the 68 and/or 78~keV lines. We therefore treat the hard X-ray lines and the gamma-ray line indepedently.

\subsection{Background modeling}
In earlier SPI analyses, the detailed background line information was neglected or used only indirectly as background tracers, obtained from cosmic-ray intensity variations or from adjacent energy bands, for example. These were separately adjusted by a set of fit parameters  for each energy bin in the measured spectra~\citep{Wang2007_60Fe,Wang2009_26Al}. To improve the spectroscopic sensitivity compared to previous SPI analysis results, we implemented a new approach for modeling the instrumental background.

We now investigate the spectroscopic signatures of nuclear reaction physics that occur in the instrument and satellite in great detail
by separately tracking the shape and intensity variations of characteristic lines in each Ge detector. First, a high-precision cumulative spectrum is used in the two energy bands, ranging from 60~keV to 85~keV, and from 1143~keV to 1175~keV, respectively, to also identify the weak instrumental line signatures. Then, a three-day integration is used to obtain sufficient statistical precision to determine spectral shape parameters of each identified background component for all detectors by fitting the spectra using a Metropolis-Hastings algorithm in a Monte Carlo Markov
chain. We then fix the shapes of the spectral templates during one orbit of the selected data set and scale the amplitude of each component for the entire Ge camera by pointing to properly trace intensity variations of each background component. Long-term investigations showed that the relative intensities of a particular spectral background feature present in one detector with respect to the mean intensity of the same feature in all detectors (detector ratios) are constant in time and thus are maximally independent of possible celestial signals~\citep{Diehl2014_SN2014J_Ni,Diehl2015_SN2014J_Co}\footnote{This is true for the time during one camera configuration. Within these 11 years of data, 4 detectors out of 19 have failed in flight, thus leading to a new camera configuration - in total, five different configurations.}. Therefore, the spectral shape and detector ratio constancy that is expected for each individual background component is imprinted onto the short-term (pointing-to-pointing) variations, while the mean amplitude (over all detectors) of a particular spectral feature is still allowed to vary to normalize the predicted background count rate.\\
In principle, each background component may have its own properties, and in particular, detector-to-detector intensity ratio, but we clearly identify physically plausible classes: Ge background lines provide a higher count rate in the inner Ge detector elements which are surrounded by Ge, while Bi background lines provide a lower count rate in the inner Ge detector elements because the BGO anticoincidence shield surrounds the Ge camera. For details see Siegert et al., in preparation (2015).

The instrument spectral response is represented as a parameterized semi-analytic function, that is, the convolution of a Gaussian with an exponential tail toward lower energies to describe the line shape. The spectral model we use in our line fitting consists of a linear continuum
\begin{equation}
C_j(E;C_{0,j},C_{1,j}) = C_{0,j} + C_{1,j} \cdot E
\label{eq:conti}
\end{equation}
and a set of instrumental lines at positions $E_{0,ij}$ , which are needed to characterize the spectrum. Each Gaussian line $i$
\begin{equation}
G_{ij}(E;A_{0,ij},E_{0,ij},\sigma_{ij}) = A_{0,ij} \cdot \exp\left(- \frac{(E-E_{0,ij})^2}{2\cdot\sigma_{ij}^2}\right)
\label{eq:line}
\end{equation}
is convolved with an exponential tail function 
\begin{equation}
T_{ij}(E;\tau_{ij}) = \frac{1}{\tau_{ij}} \exp\left(-\frac{\tau_{ij}}{E}\right)
\label{eq:tail}
\end{equation}
that describes the impact of cosmic radiation that gradually deteriorates the charge collection efficiency of each detector $j$. The raw spectra of SPI are background dominated; the instrumental background is more than $99\%$ of the total, with a small contribution from a celestial signal. Fitting the raw spectra for each detector and a three-day cumulative sample results in a very precise spectrum of what is observed from instrumental background gamma-ray lines and continuum. This "continuum" includes the instrumental background continuum ($>99\%$) as well as any possible continuum emission from celestial sources. We are mainly interested in the emission of gamma-ray lines; therefore, this approach largely also suppresses the continuum from celestial sources, and any resulting offset from zero in the finally derived spectra can be considered as a measure of systematics of this continuum treatment. For each energy bin, a discrete background detector pattern (in time and position in the sky) is predicted, which serves as the only background component $B_{jk}$ in Eq.~(\ref{eq:model-fit}).
The degradation of Ge detectors from cosmic-ray irradiation and their restoration in annealings results in a time-variable width and asymmetry of the spectral response. This variation dominates all other spectral changes and is found consistent across the SPI energy range.

\section{Results and discussion}
\label{sec:results}
\subsection{\Ti gamma-ray line measurements}

Our analysis shows the expected signature of a (Gaussian-shaped) line at the expected energies for the two independent spectra. The significances of the lines are 2.2~$\sigma$ (1157~keV) and 3.1~$\sigma$ (78~keV), with a combined \Ti detection significance of $3.8~\sigma$. Unfortunately, systematic effects due to strong background features below 70 keV cannot be reliably modeled; therefore no constraint on the 68 keV line could be obtained. The mean reduced $\chi^2$ of the derived fit per energy bin is 1.06 ($\chi^2=98241$, d.o.f. $=92517$), suggesting an additional systematic uncertainty of  \about~6\%. This is about eight times lower than the statistical uncertainties.

\begin{figure}[htbp]
  \centering
  \includegraphics[width=\linewidth]{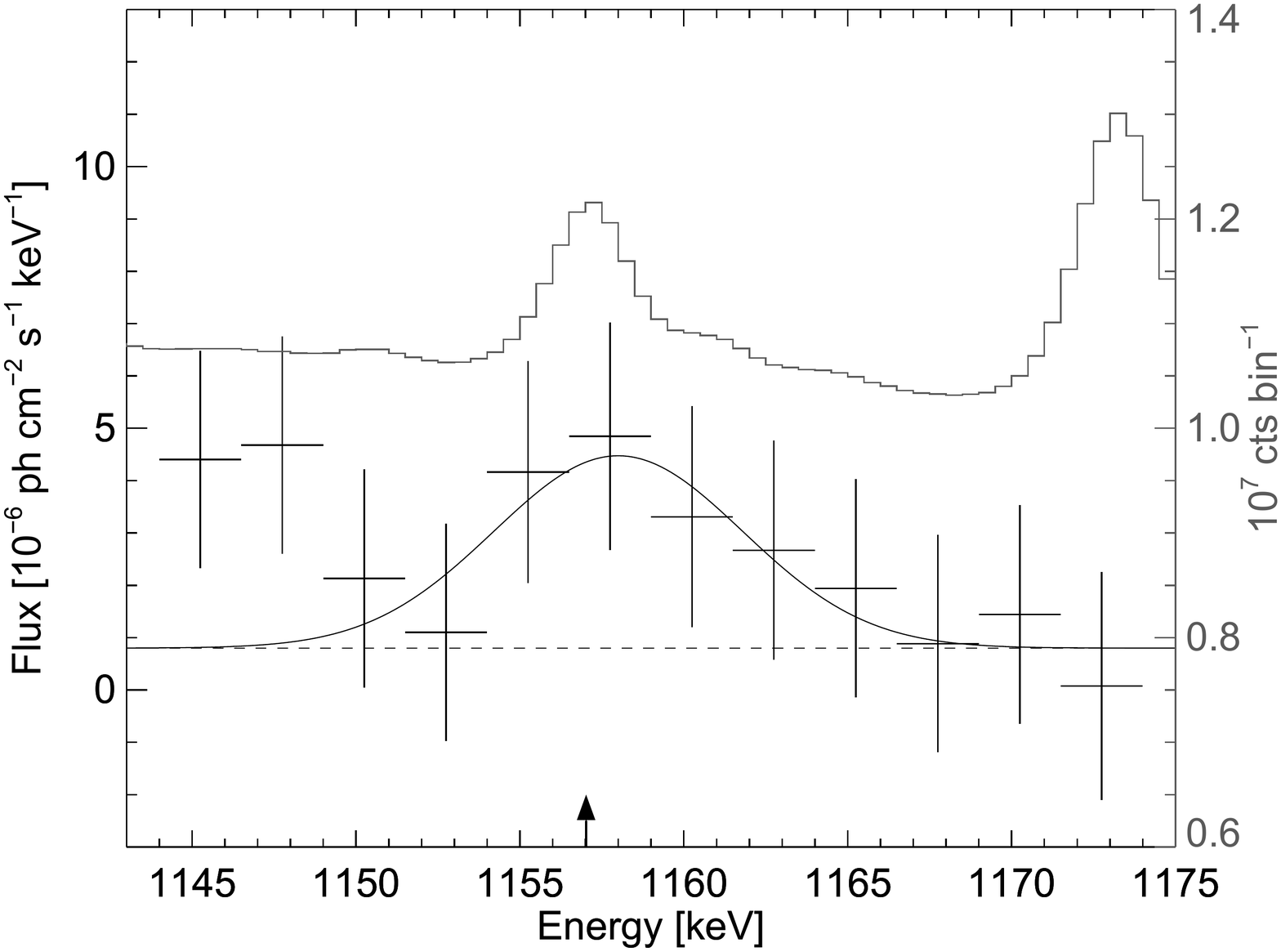}
  \caption{Spectrum obtained from a source at the position of Cas A at 1157 keV (black crosses, 2.5~keV energy bins). The spectrum has been fitted by a constant plus a Gaussian that is centered at $(1158\pm3.6)~\mathrm{keV}$ and broadened $(8.94\pm1.42)~\mathrm{keV}$ (FWHM). The derived expansion velocity of the \Ti ejecta is $(2200\pm400)$~$\mathrm{km~s^{-1}}$. The measured flux is $(3.5 \pm 1.2)~10^{-5}$~ph~\cms , which can be converted to an observable \Ti mass of $(2.4\pm0.9)~10^{-4}$~\Msol at the time of the explosion. The cumulative long-time spectrum (mainly background) is shown as a gray histogram. The laboratory-determined energy of the excited \Ca line is marked with an arrow.}
  \label{fig:CasA_spec_1157}
\end{figure}

\begin{figure}[!htbp]
  \centering
  \includegraphics[width=\linewidth]{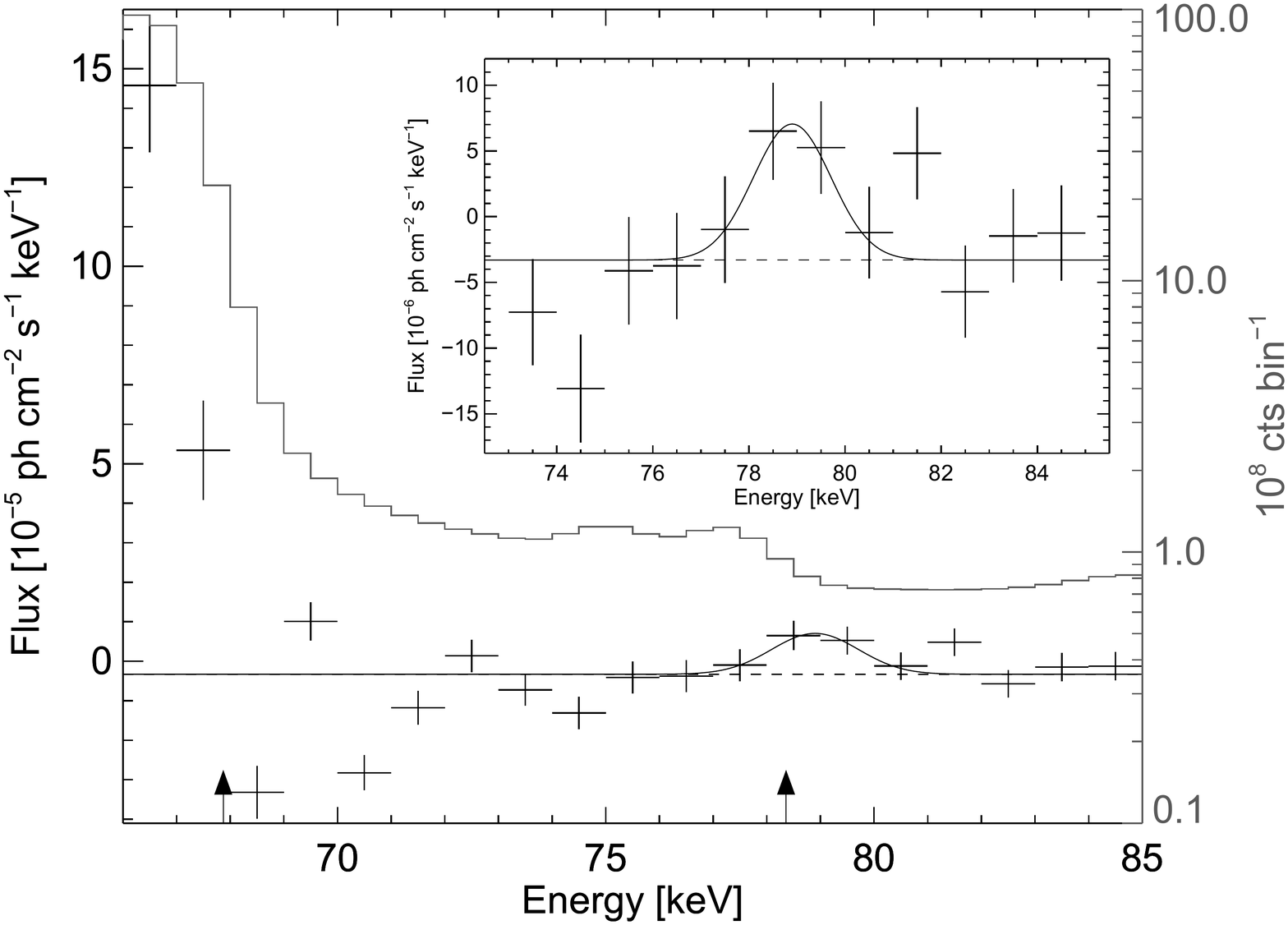}
  \caption{Same as Fig.~\ref{fig:CasA_spec_1157}, but at 78 keV (1.5~keV energy bins). The best-fit values are $(78.9\pm1.5)~\mathrm{keV}$ for the centroid, $(1.9\pm~0.6)~\mathrm{keV}$ for the width (FWHM), yielding an expansion velocity of $(4300\pm1600)~\mathrm{km~s^{-1}}$, and $(2.1 \pm 0.4)~10^{-5}$~ph~\cms for the flux, which is equivalent to $(1.5\pm0.4)~10^{-4}$~\Msol of \Ti. The raw count spectrum (gray) increases by two orders of magnitude below 70~keV with respect to 80~keV, which introduces strong systematic effects.}
  \label{fig:CasA_spec_78}
\end{figure}

In Figs.~\ref{fig:CasA_spec_1157} and~\ref{fig:CasA_spec_78}
we show the derived spectra around the 1157~keV line and the 78~keV line. The spectrum in Fig.~\ref{fig:CasA_spec_1157} is well represented by a constant plus a Gaussian that is centered at $(1158\pm3.6)~\mathrm{keV}$ and broadened with respect to the instrumental resolution.
We obtain a total flux of $F_{1157} = (3.5 \pm 1.2)~10^{-5}$~ph~\cms from this fit. This can be translated into an observable \Ti mass at the time of the explosion ($t=0$) of $(2.4\pm0.9)~10^{-4}$~\Msol, by
\begin{equation}
M_{^{44}Ti}(t=0) \cong 4 \pi d^2 F_{1157}(t) \cdot 44 m_u \cdot \tau_{^{44}Ti} \cdot \exp(t/\tau_{^{44}Ti}) / W_{1157},
\label{eq:mass_cal}
\end{equation}
taking into account that the characteristic lifetime of \Sc (\SI{5.73}{\hour}) is much shorter than of \Ti (\SI{86}{\year}). In Eq. (\ref{eq:mass_cal}), $d = 3.4^{+0.3}_{-0.1}$~kpc is the distance to the SNR, $t = (337.2\pm6.4)~\mathrm{a}$ its age, incorporating the eleven-year observation time span, $F_{1157}(t)$ the measured flux in the 1157~keV line today, $m_u$ the atomic mass unit, $\tau_{^{44}Ti} = (86\pm0.5)~\mathrm{a}$ the characteristic lifetime of the \Ti decay, and $W_{1157} = 99.9\%$ the respective branching ratio. 

The uncertainty on this \Ti mass value is mainly due to the distance to Cas A and the measured flux.
In addition, the \Ti half-life time has an impact on the derived mass; its uncertainty is taken into account above. Because statistical fluctuations in the measured spectra are large, the fit and the resulting mass might suffer from systematic effects in the assumed spectral shape (constant plus Gaussian); we estimate this effect by determining an upper limit of the constant offset, which would then reduce the derived mass by about 33\%.

The high-energy \Ti decay line has a FWHM of $(8.94\pm1.42)~\mathrm{keV}$ and is thus broadened with respect to the instrumental resolution at 1.157 MeV of \about~$2.4$~keV~\citep{Attie2003_SPI}. This FWHM can be converted to an estimate of the expansion velocity of the \Ti ejecta of $(2200\pm400)$~$\mathrm{km~s^{-1}}$. This value is significantly lower than values reported in recent papers~\citep{Renaud2006_CasA,Grefenstette2014_CasA}. The line centroid is found as $(1158.0\pm3.6)~\mathrm{keV}$ and thus agrees with the laboratory-determined energy, suggesting only minor bulk average motion of \Ti in Cas A relative to the solar system, ($v_{Bulk} = (-250\pm1000)~\mathrm{km~s^{-1}}$).

\begin{figure}
  \centering
  \includegraphics[width=\linewidth]{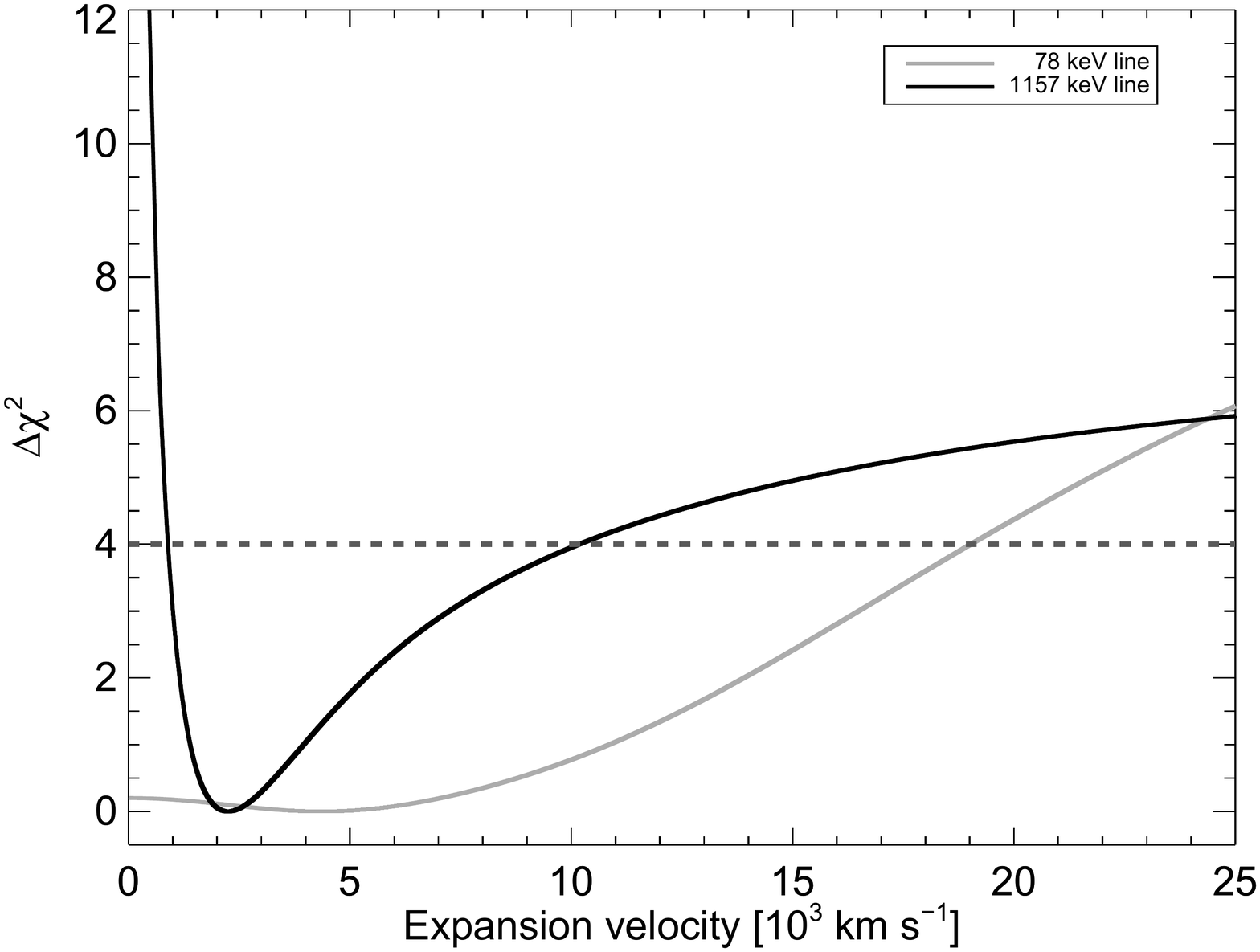}
  \caption{$\Delta\chi^2$ vs. expansion velocity for the two detected lines, individually. The values are calculated assuming the best-fit values in each case, only letting the Gaussian width vary. Because of the  higher resolving power of SPI at 1 MeV with respect to lower energies, the value for the expansion velocity derived form the high-energy line (solid black) is more constrained than that from the low-energy line (solid gray). For comparison, the 2$\sigma$ limits are indicated by the thick dashed line.}
  \label{fig:upper_limit}
\end{figure}

For the 78 keV line (Fig.~\ref{fig:CasA_spec_78}), we obtain a line flux of $F_{78} = (2.1 \pm 0.4)~10^{-5}$~ph~\cms. Using Eq.~(\ref{eq:mass_cal}) and replacing $W_{1157}$ with $W_{78} = 96\%$, $F_{78}$ is converted to a \Ti mass of $(1.5\pm0.4)~10^{-4}$~\Msol. Within uncertainties, the two derived \Ti masses are consistent. 
The 78~keV line is not significantly shifted either, confirming absence of average bulk motion. 
The measured Doppler broadening of $(1.9\pm~0.6)~\mathrm{keV}$ (FWHM) yields an expansion velocity of $(4300\pm1600)~\mathrm{km~s^{-1}}$ for the \Ti ejecta. This value is consistent with the result from the 1157~keV line and also with the value measured by NuStar. The apparent differences between values derived for the 78 and 1157~keV lines reflect the energy dependence of the resolving power of the SPI instrument: While at $78.4~\mathrm{keV}$, the resolving power of the spectrometer is $R_{78}=78.4/1.6=49$, it is $R_{1157}=1157/2.4=482$ at $1157~\mathrm{keV}$, meaning
that SPI derives better line width constraints at higher energies; this adds to the linear energy dependence of the Doppler effect.
Figure~\ref{fig:upper_limit} shows the 2$\sigma$ upper limit on the expansion velocity assuming the best-fit values for flux and centroid for the two detected lines. While the constraint from the 78~keV line is rather flat with a shallow minimum (gray line in the figure), the 1157~keV line (black line) provides a well-defined minimum or constraint.

Representing the line shapes by Gaussians is an empirical approach. Alternatively, a physical model can be adapted: Assuming a homogeneously expanding thin shell where the shell radius is much larger than the distance to the observer, the expected ideal line shape can be described by a tophat function
\begin{equation}
S(E;F_{0},E_{0},\dot{R}) = F_{0} \cdot \Theta\left(\left|\dot{R}\right| - \left|c\left(\frac{E}{E_{0}}-1\right)\right|\right) \cdot \frac{c}{2\dot{R}E_{0}},
\label{eq:thin_shell}
\end{equation}
where $F_{0}$ is the measured flux, $E_{0}$ the bulk motion, $\dot{R}$ the expansion velocity, and $c$ the speed of light~\citep{Kretschmer2011_PhD}. This line shape has to be convolved with the spectral response function of SPI (see Eqs.~(\ref{eq:line}) and~(\ref{eq:tail})). Fitting the spectra using Eq.~(\ref{eq:thin_shell}) leads to lower velocities in both cases, that is, $(3900\pm800)~\mathrm{km~s^{-1}}$ (78~keV) and $(1500\pm400)~\mathrm{km~s^{-1}}$ (1157~keV), respectively. The measured expansion velocities, however, are similar to those from a Gaussian fit, and systematic differences are smaller than uncertainties from statistics.

When the results of~\citet{The1996_CasA} obtained with OSSE on CGRO with the COMPTEL result by~\citet{Iyudin1997_CasA}, the BeppoSAX result~\citep{Vink2001_CasA}, the INTEGRAL/IBIS result~\citep{Renaud2006_CasA}, and the NuStar result~\citep{Grefenstette2014_CasA} are combined with our work through a weighted mean (weighted by the inverse variance of each measurement), the  \Ti mass is $(1.37\pm0.19)~10^{-4}$~\Msol. 
\begin{figure}
  \centering
  \includegraphics[width=\linewidth]{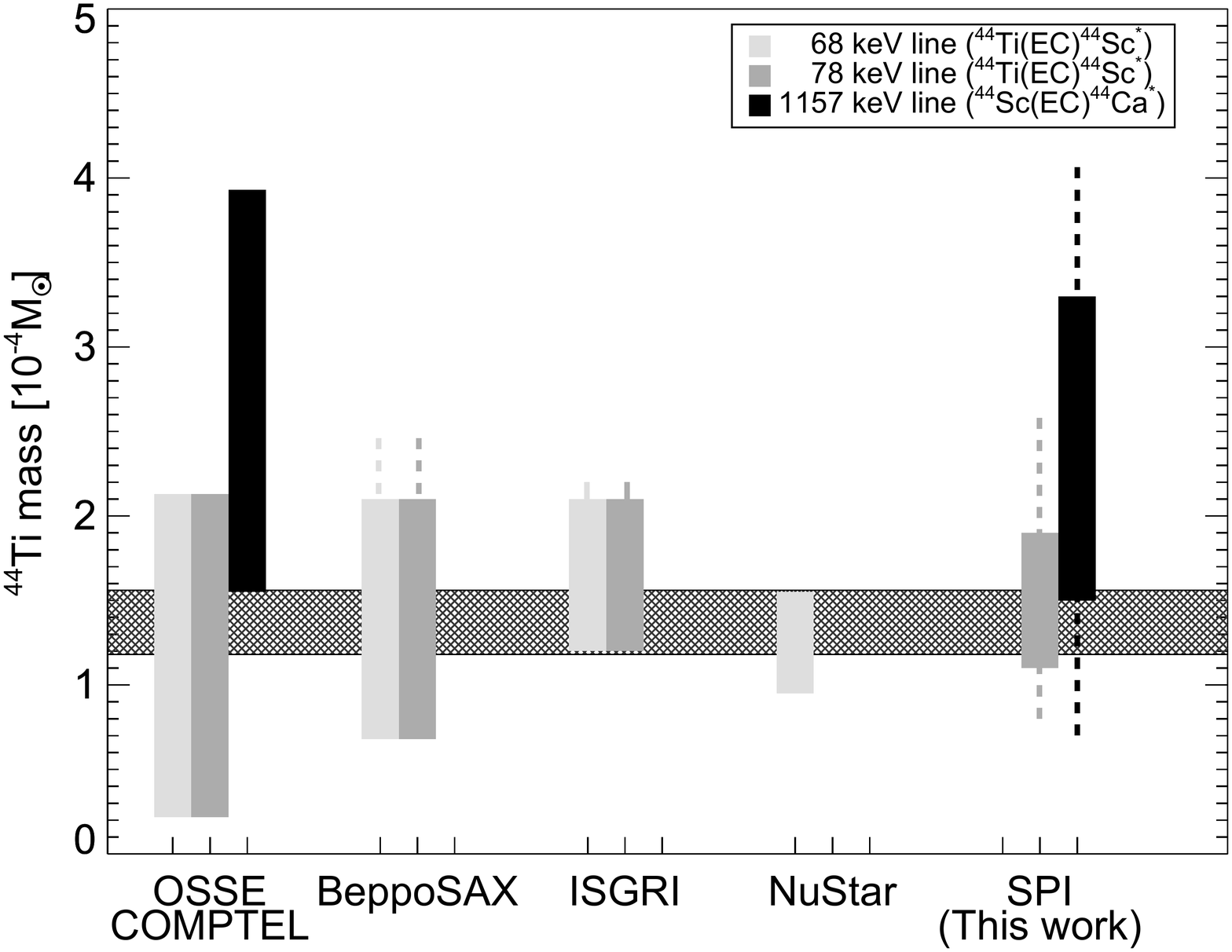}
  \caption{Derived \Ti masses of different instruments. All data bars are drawn from the measurements with only the given instrument (no combinations), including the uncertainties in half-life time and explosion date. The OSSE and COMPTEL values are corrected for the distance to Cas A. Upper limits (if given) are shown as dashed lines. The yields measured by COMPTEL and SPI using the 1157 keV line are systematically higher than values using the low-energy lines. The calculated weighted mean of all shown measurements is illustrated as the hatched area and is $(1.37\pm0.19)~10^{-4}$~\Msol.}
  \label{fig:masses}
\end{figure}

\subsection{Flux consistency check}
\begin{figure}
  \centering
  \includegraphics[width=\linewidth]{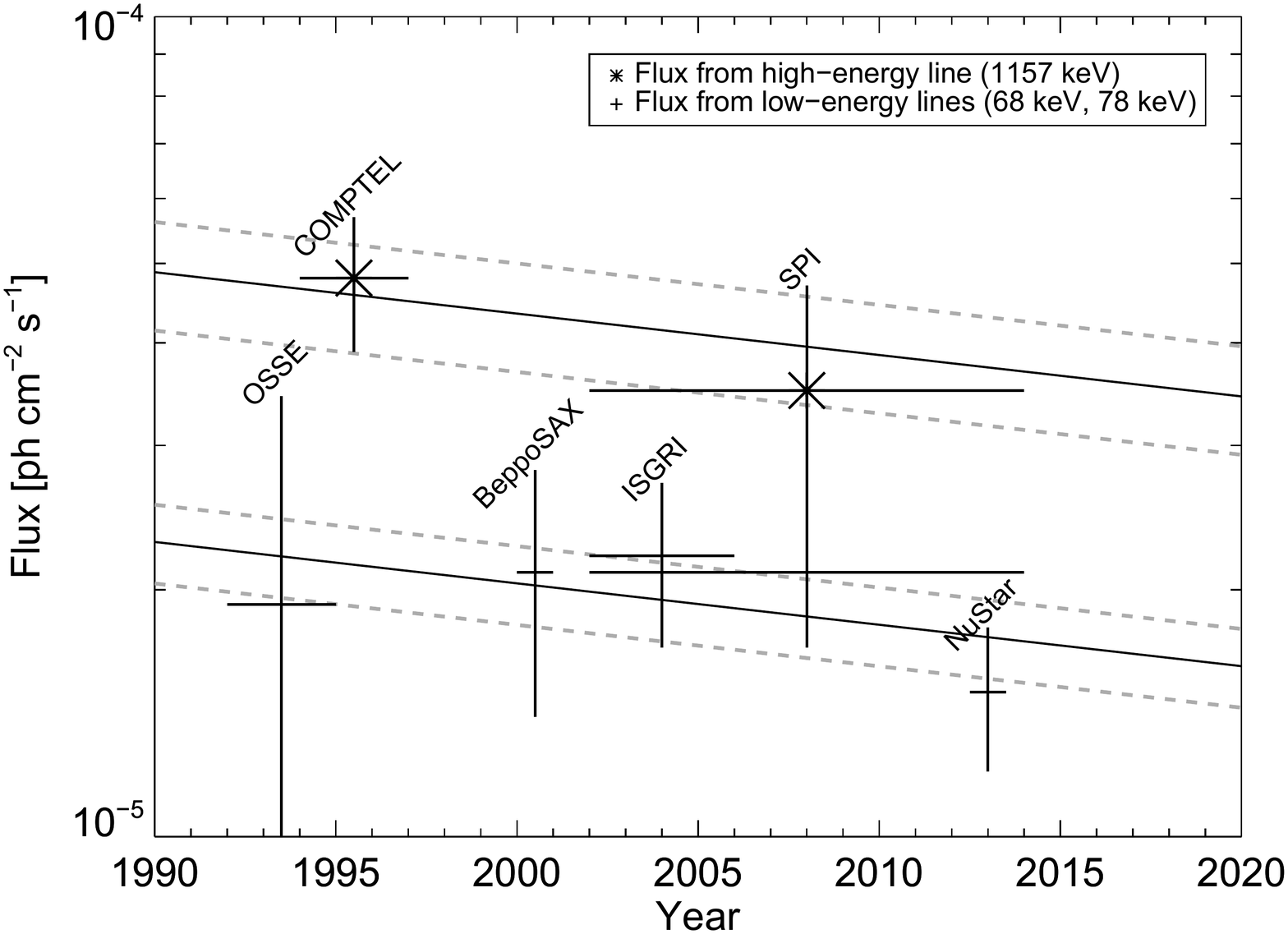}
  \caption{Compilation of $\gamma$-ray flux measurements in Cas A from different instruments during the years 1992 to 2014 (see text). To each family of lines (low- and high-energy), an exponential decay function has been fitted, fixing the half-life time of \Ti to 86~a and the explosion date to AD 1671 (solid lines). The 1$\sigma$ uncertainties are separately shown by dashed gray lines for each fit. The discrepancy can be expressed as a constant flux level of $(2.29 \pm 0.62)~10^{-5}$~ph~\cms.}
  \label{fig:hilo_fluxes}
\end{figure}

We note that the high-energy line appears to yield a systematically higher \Ti mass, both from our analysis and in~\citet{Iyudin1994_CasA} and~\citet{Iyudin1997_CasA}, with respect to the hard X-ray line based \Ti mass determinations. Our derived \Ti mass from the 78~keV line is consistent with other published studies that measured the low-energy lines. The derived \Ti mass per instrument is shown in Fig.~\ref{fig:masses} together with the weighted mean of all measurements (hatched area).  
This systematic difference in \Ti flux values becomes even more obvious when the measurements of the \Sc and \Ca lines, taken with different instruments in the past two decades, are compared and plotted on a time axis (Fig.~\ref{fig:hilo_fluxes}).
The values in Fig.~\ref{fig:hilo_fluxes} are taken from~\citet[][Table 3]{The1996_CasA} with OSSE,~\citet{Iyudin1997_CasA} with COMPTEL,~\citet[][Table 1]{Vink2001_CasA} with BeppoSAX,~\citet[][Table 1]{Renaud2006_CasA} with IBIS, this work with SPI, and~\citet[][Table ED2]{Grefenstette2014_CasA} with NuStar (from left to right). These values only consider the value measured with the mentioned instrument itself, hence no double-counting of flux values.
The 1157~keV flux values (\citet{Iyudin1997_CasA} and our study) are found to be significantly higher than the low-energy line flux measurements.

We investigated this systematic by treating the measurements of the hard X-ray lines independently of the 1157~keV gamma-ray line. We performed an F-test to check whether the derived spectra can be represented by only one flux value (case I) or if the preference is for two values (case II). The corresponding $\chi^{2}$ values are $\chi^{2}_{I} = 15.14$ (16 dof), and $\chi^{2}_{II} = 12.27$ (15 dof), respectively. The F-value of this test is $F=3.51,$ for which the F-statistic gives a probability of $\sim1\%$, corresponding to $2.5~\sigma$, that the two spectra cannot be represented by the same flux value.\\
Alternatively, we fixed the explosion date to AD 1671 and the characteristic lifetime of \Ti to 86 years and then fit an exponential decay function to only the flux measurements of the low-energy lines. This yields a \Ti mass of $(1.28\pm0.14)~10^{-4}$~\Msol. The same fitting procedure applied to the 1157~keV gamma-ray line flux measurements (COMPTEL and SPI) results in a \Ti mass of $(2.72\pm0.43)~10^{-4}$~\Msol. Using the first value as an estimate for the real observable \Ti mass seen to decay, that
is, uncontaminated by possible secondary processes that might mimic additional \Ti, we evaluated an additional flux level of $(2.29 \pm 0.62)~10^{-5}$~ph~\cms by fitting the high-energy measurements with a constrained \Ti mass. This discrepancy has a statistical significance of 3~$\sigma$.

About 340 years after the explosion, the expected flux ratio $F_{78}/F_{1157}$ is $\sim1$, the measured flux ratio is $\sim0.5$. This would imply an efficiency (number of photons per decay) error of 50\% and is excluded by several measurements~\citep{Norman1998_44Ti,Wietfeldt1999_44Ti,Ahmad2006_44Ti}.

\subsection{Nuclear de-excitation?}
We speculate that this discrepancy may be due to an additional nuclear de-excitation component originating from accelerated particles colliding with $^{44}\mathrm{Ca}^{*}$ in the remnant and swept up interstellar medium. CRs are thought to be accelerated in SNRs by diffusive shock acceleration~\citep{Baade1934_DSA}, in particular in Cas A~\citep{Berezhko2004_CasA,Aharonian2001_CasA}. Only the 1157~keV line from excited \Ca will have an additional contribution from nuclear excitation (see Sect.~\ref{subsec:treatment}). The additional gamma-ray line flux originating from the isotope \Ca can be estimated by
\begin{equation}
F_{\gamma} = (4 \pi d^2)^{-1} \cdot n(^{44}\mathrm{Ca}) \cdot \int Q_{P}(p) \sigma(p) v(p) dp\mathrm{,}
\label{eq:deex_flux}
\end{equation}
where d is the distance to Cas A, $n(^{44}\mathrm{Ca})$ the mean density of \Ca atoms in the medium where the interaction takes place, $Q_{P}(p)$ the proton acceleration spectrum\footnote{The proton acceleration spectrum, $Q_{P}(p) \propto p^{-2.3}$, was chosen following~\citet{Summa2011_CasA} to meet the FERMI-LAT constraints on the energy content of the cosmic rays that were measured by~\citet{Abdo2010_CasA}, about 3$-$4$~10^{49}~\mathrm{erg}$.}, $\sigma(p)$ the cross section of the reaction $^{44}\mathrm{Ca}(p,p')^{44}\mathrm{Ca}^{*}$, and $v(p)$ the velocity of accelerated protons. The product of $n(^{44}\mathrm{Ca})$ by the integral, $I(^{44}\mathrm{Ca})$ in Eq.~(\ref{eq:deex_flux}) describes the collision frequency and includes the present $^{44}\mathrm{Ca}$ abundance in the interaction region~\citep{Ramaty1979_gamma,Summa2011_CasA}. 

With Eq.~(\ref{eq:deex_flux}), the  \Ca density required in front of the shock to produce this gamma-ray flux can be estimated. The mean cross sections of the reactions $^{12}\mathrm{C}(p,p')^{12}\mathrm{C}^{*}$ by~\citet{Ramaty1979_gamma}, and $^{44}\mathrm{Ca}(p,p')^{44}\mathrm{Ca}^{*}$ by~\citet{Mitchell1982_cs} have the same order of magnitude. With the solar abundance ratio $^{44}\mathrm{Ca}/^{12}\mathrm{C} = \frac{1311}{7.001\cdot10^{6}} \approx 2 \cdot 10^{-4} $ by~\citet{Lodders2003_abundances}, the integral $I(^{44}\mathrm{Ca})$ can be estimated to be $2\cdot10^{-4}I(^{12}\mathrm{C})$. 
$I(^{12}\mathrm{C})$ was calculated from the values given in~\citet{Summa2011_CasA}. The resulting \Ca density is about $10^5$ - $10^6~\mathrm{cm^{-3}}$. This value is very high compared to densities of the interstellar medium, but may reflect the swept-up density in front of the forward shock (``snow plough''). If the measurements of presolar grains by~\citet{Nittler1996_Dust},~\citet{Clayton1997_Dust}, and~\citet{Hoppe2000_Dust} are taken into account, which suggest that the ratio $^{44}\mathrm{Ca}/^{40}\mathrm{Ca}$ can be more than 100 times higher than solar, the forward-shock \Ca density is scaled down correspondingly  to $10^3$ - $10^4~\mathrm{cm^{-3}}$. Although this density range appears plausible for optical knots and might even be consistent with expectations~\citep{Fesen2006_CasA,Docenko2010_CasA}, the estimated values appear extreme: Assuming a volume of the swept-up material near the Cas A SNR of $\sim10^{56}~\mathrm{cm^{3}}$, and a density of $~\sim10^3~\mathrm{cm^{-3}}$, the estimated \Ca mass would be of the order $10^3$~\Msol. Therefore, the ejecta from Cas A alone cannot explain this mass; adding the swept-up interstellar material cannot explain the additional flux either. The morphology of the swept-up material possibly differs from the expectations, or the physics of nuclear excitation may not be properly treated in our simple estimate. It would be interesting to investigate the nuclear de-excitation lines from $^{12}\mathrm{C}$ and $^{16}\mathrm{O}$ in Cas A because these lines are expected to be more luminous.

Additional support for our speculated nuclear de-excitation component is the measured Doppler velocity of the 1157~keV line: It is lower than the one derived from the 78~keV line.  While the emission originating in the decay of \Ti to \Sc (78~keV line) follows the kinematics of the expanding SNR, the 1157~keV line emission might have two components, one resembling the kinematics (\about$5000~\mathrm{km~s^{-1}}$) from the \Sc decay to \Ca, and another one incorporating the nuclear de-excitation of $^{44}\mathrm{Ca}^{*}$ in the swept-up material in front of the shock at zero velocity. However, we do not know where the interaction takes place and can also assume that a measurement of the 1157 keV line samples another volume element of Cas A than the 78 keV line. \citet{Milisavljevic2015_CasA} illustrated the bubble-like behavior of the interior of Cas A that naturally shows velocities of about $2000~\mathrm{km~s^{-1}}$ near the expansion center and up to $5000~\mathrm{km~s^{-1}}$ farther outside.

\section{Conclusions and discussion}
\label{sect-conclusions}

Revisiting the 340 year old supernova remnant Cassiopeia A with eleven years of data from the spectrometer SPI on INTEGRAL, we detected two gamma-ray lines originating in the \Ti decay chain $^{44}$Ti$\rightarrow$$^{44}$Sc$\rightarrow$$^{44}$Ca. 
One of the low-energy lines from the process $^{44}$Ti$\rightarrow$$^{44}$Sc at 78~keV was detected, while the adjacent line at 68~keV coincides with a major and difficult instrumental-background feature and cannot be measured. The 78~keV line parameters are consistent with other measurements constraining the \Ti mass and kinematics in the supernova remnant, ($(1.5\pm0.4)~10^{-4}$~\Msol, and $v_{exp} $\about$ (4300\pm1600)~\mathrm{km~s^{-1}}$). The second decay step in the \Ti decay chain, $^{44}$Sc$\rightarrow$$^{44}$Ca, occurs only a few hours after the \Ti decay and produces characteristic 1157~keV gamma-rays that were also detected and analyzed independently. The \Ti mass from the 1157~keV line is $(2.4\pm0.9)~10^{-4}$~\Msol, while Doppler broadening here implies an expansion velocity of $(2200\pm400)~\mathrm{km~s^{-1}}$. 

The kinematic information of the Cas A SNR by measuring emission lines is not unique. In particular, fast-moving O-rich knots at velocities of around 8000 km~s$^{-1}$ have been found~\citep{Fesen2006_CasA}. From Chandra measurements, characteristic X-ray lines of Si and Fe have been mapped. An inversion of the classical onion-shell structure had been inferred, as Fe X-ray emission clumps have been found outside the Si X-ray emitting regions more central to the remnant~\citep{Hwang2004_CasA}. From the NuStar image and the more central location of \Ti clumps~\citep{Grefenstette2014_CasA}, this inversion hypothesis may be recognized as a misinterpretation of the Fe line X-ray emission, which is driven by the ionization state of remnant gas; Fe in the central region thus may escape X-ray line detection. The forward shock reached a diameter of 5 arcmin and appears to currently move at 5000~km~s$^{-1}$, while the X-ray bright regions predominantly are beyond the inner 2~arcmin, and in the jet regions at radial distances above 3~arcmin from the central compact object~\citep{Hwang2004_CasA}. Thus the reverse shock has progressed about half into the remnant (inner diameter $\sim$2--3~arcmin) and has not yet ionized the inner regions  (see also~\citep{Milisavljevic2015_CasA} and~\citep{Grefenstette2015_CasA} for discussions in NIR, and X-ray continuum wavelengths, respectively).

\Ti nucleosynthesis yields in core-collapse supernovae is one of the questions we wish to address; they could range up to several 10$^{-5}$~\Msol~\citep{Timmes1996_44Ti,The2006_44Ti,Magkotsios2010_44Ti}, while for thermonuclear supernovae (SN Ia) only a rare subtype is considered a plausible \Ti source, but then also with yields of up to several 10$^{-5}$~\Msol~\citep{Woosley1994_SNIa}, see also~\citet{The2006_44Ti}. 
Nuclear burning in supernova explosions occurs at high densities and temperatures, and thus those sites are closest to conditions of equilibrium burning, where nucleons find their most stable arrangement within atomic nuclei. The binding energy of nucleons is maximized for \Ni, and nuclear statistical equilibrium (NSE) appears to be a good description of plasma conditions for the brief moment of the supernova explosion when nucleosynthesis occurs. Here all nuclear reactions except weak-force reactions are in thermal balance. As the explosion site is diluted and cools, the first nuclear reactions falling out of such equilibrium are three-body reactions such as the 3$\alpha$ conversion of helium to carbon. Here, the conditions are better described by an overabundance of $\alpha$ particles over NSE. This is called \emph{$\alpha$-\textup{rich freeze-out}}, and nucleosynthesis productions emerging from these conditions are characterized by overabundances of nuclei that are multiples of $\alpha$ particles resulting from successive captures of $\alpha$s on nuclei~\citep{Bodansky1968_QSE,Woosley1973_freezeout}.
$^{40}$Ca and \Ti are among the most-massive of such products; note that \Ni  is an $\alpha$-multiple nucleus as well. It has been considered plausible, therefore, that supernovae that produce major amounts of \Ni are also sources of significant amounts of \Ti.

When the measurements taken by various instruments during the last 20 years are combined, the \Ti mass seen to decay is $(1.37\pm0.19)~10^{-4}$~\Msol.
But we find a significant difference ($3~\sigma$) in flux, resulting \Ti mass, and velocity from the measurements of the \Ca versus $^{44}$Sc lines (high- and low-energy lines, respectively). We argue that the low-energy line measurements (68~keV and/or 78~keV) reflect the true \Ti content and kinematics, and an additional component with a flux level of $(2.29 \pm 0.62)~10^{-5}$~ph~\cms is found in the 1157~keV line flux. We speculate that this excess is due to the nuclear de-excitation of $^{44}$Ca$^{*}$ in the remnant and swept-up interstellar medium, excited by cosmic rays that are expected to be accelerated in the young remnant by diffusive shock acceleration.

%
\begin{acknowledgements}
  This research was supported by the German DFG cluster of excellence
  ``Origin and Structure of the Universe''. The INTEGRAL/SPI project
  has been completed under the responsibility and leadership of CNES;
  we are grateful to ASI, CEA, CNES, DLR, ESA, INTA, NASA and OSTC for
  support of this ESA space science mission.
\end{acknowledgements}

%

%

\end{document}